\documentclass[12pt]{article}

\usepackage{graphicx}
\usepackage{color}

\begin{document}
 \title{Superconducting Fluctuations in 
 a Tight Binding Bandstructure}
 \author{L. Coffey,\\
 Physics Department,\\ 
 Illinois Institute of Technology,\\
 Chicago, Illinois 60616}
\maketitle
\begin{center}
Normal state superconducting fluctuations are calculated for the case of a tight 
binding bandstructure. The resulting electronic self energy $\Sigma( \vec{p},
\omega)$ and spectral weight 
$A( \vec{p}, \omega)$ are anisotropic on the Fermi surface. 
For certain values of the chemical potential $\mu$, the onset of a pseudogap is
present in $A( \vec{p}, \omega)$ for Fermi surface momenta near the ($\pi$,0) 
point in the 1st Brillouin Zone, and absent for momenta closer to the zone diagonal. 
The pseudogap in the normal state of high temperature superconductors shows similar behavior.
\end{center}

\newpage

One of the well known predicted effects of superconducting fluctuations in the normal state above the transition temperature $T_{C}$ is
a reduction in the density of states about the Fermi energy, which can be calculated \cite{Ab} using a self energy $\Sigma(\vec{p}, \omega)$  containing a Boson Green function for the fluctuations known as the fluctuation propagator  $L(q,\omega)$ \cite{Russ}.\\
\\
The goal of this paper is to calculate $L(q,\omega)$ with a tight binding bandstructure, 
and to investigate the properties of the resulting fluctuation electronic self-energy 
$\Sigma( \vec{p},\omega)$ and spectral weight $A( \vec{p}, \omega)$. \\
\\
The calculations in this work produce a $\Sigma(\vec{p},\omega)$ and an $A( \vec{p}, \omega )$ that are strongly
anisotropic on the Fermi surface when the value of the chemical potential $\mu$ in the tight binding bandstructure is chosen to be close to the
energy of the Van Hove singularity in the density of states. For such a choice of $\mu$, the 
magnitude of  $\Sigma( \vec{p},\omega)$ is large for Fermi surface momenta near the $(\pi,0)$ point in the 1$^{st}$ Brillouin Zone (BZ), from where the Van Hove singularity in the density of states originates, and is small for  Fermi surface momenta closer to the zone diagonal of the BZ. 
A reduction in spectral weight, or a pseudogap, appears in the electronic spectral weight $A(\vec{p}, \omega)$ for Fermi surface momenta near $(\pi,0)$, and is absent for Fermi surface momenta closer to the BZ zone diagonal 
direction. This behavior is essentially a result of the {\em hot spot} role of momenta near the
$(\pi,0)$ point in determining tight binding band electronic properties.\\ 
\\
In this work, $L(q,\omega)$ is calculated using a momentum independent attractive
interaction between electrons, with the Fermi surface anisotropy in 
$\Sigma(\vec{p},\omega)$ and $A( \vec{p}, \omega )$ the result of the tight 
binding bandstructure alone.
The use of a momentum independent electron-electron interaction in 
calculating $L(q,\omega)$ implies s-wave superconductivity.
A momentum dependent normal state interaction $V(\vec{p},\vec{p^{'}})$ would be required
to investigate fluctuations for the onset of d-wave superconductivity.
However, even with the limitation of a momentum independent interaction, the results
of this work may still provide useful information on the role of band structure in
superconducting fluctuations in cases such as high temperature cuprates.\\ 
\\
This work is motivated by the properties of the pseudogap measured in angle resolved photoemission spectroscopy
(ARPES) in the high $T_{C}$ cuprates \cite{Review15}. The origin of this pseudogap is an active area of investigation.
One of its measured properties is that it is non-zero at the Fermi surface antinode of the cuprate d-wave superconducting gap ($(\pi,0)$ point), and zero along an arc of momenta on the Fermi surface about the superconducting gap nodal point on the BZ zone diagonal. \\
\\
The role of fluctuations in generating pseudogaps has been the subject of extensive
theoretical study \cite{BJ,Per,Spine,Har}, as well as their effect on high $T_{C}$ density of states \cite{Cast} 

\newpage

{\bf Theoretical Formalism}\\
\\
The fluctuation propagator $L(q,\omega)$ is defined as
\begin{equation}
L(q, \omega) \; = \; \frac{c}{1 \; - \; c P(q,\omega)}
\end{equation}
where $c$ is the superconducting coupling constant whose value is chosen to yield a divergence
in $L(q=0,\omega=0)$ at a chosen transition temperature $T_{C}$. \\

The particle-particle propagator $P(q, \omega)$ is given by
\begin{equation}
P(q, \omega) \; = \; 
\int \frac{d^{2} k}{(2 \pi)^{2}} 
\frac{1 - n_{F} (\varepsilon_{k}) - n_{F} (\varepsilon_{q-k})}
{\omega -\varepsilon_{k} - \varepsilon_{q-k} + i \delta}
\end{equation} 
incorporating a tight binding bandstructure defined by
\begin{equation}
\varepsilon_{k} \; =\; -2t( {\rm cos}(k_{x})+{\rm cos}(k_{y}))+4t^{'} {\rm cos}
(k_{x}) {\rm cos}(k_{y})-
\mu
\end{equation}
$4t^{'}$ is chosen to be $1.4t$.\\

The electronic self energy $\Sigma(\vec{p}, \omega)$ due to fluctuations is 
\begin{equation}
\Sigma(\vec{p}, \omega) \; = \; \int \frac{d^{2} q}{(2 \pi)^{2}}  \int 
\frac{d \omega^{'}}{\pi} 
{\rm Im} L(q, \omega^{'})
\frac{[ n_{B}( \omega^{'}) + n_{F}( \varepsilon_{q-p})] }{ 
\omega^{'} - \omega -\varepsilon_{q-p} - i \delta }
\end{equation}
\\
with $n_{B}( \omega)$ and $n_{F}(\omega)$ denoting the Bose-Einstein and Fermi-Dirac distributions, respectively.\\

Finally, the electronic spectral weight is calculated from
\begin{equation}
A( \vec{p}, \omega )\; = \; - \frac{1}{\pi} \frac{ {\rm Im} \Sigma(\vec{p}, \omega) }
{(\omega - \varepsilon_{p} - {\rm Re} \Sigma(\vec{p}, \omega ))^{2} + ({\rm Im} 
\Sigma(\vec{p}, \omega))^{2} }
\end{equation}
\\
\\
{\bf Results}\\
\\
Figures 1 to 5 show a selection of typical results for $\Sigma(\vec{p}, \omega)$ 
and $A( \vec{p}, \omega )$.\\
\\
The real and imaginary components of the self energy $\Sigma(\vec{p}, \omega)$ are 
shown in figure 1
for the case $\mu=-1.3t$, $T_{C}=0.2t$ and temperature $T=0.22t$. The superconducting coupling constant in equation (1) is
$c=-1.7498t$.  
Results for two Fermi surface points are displayed: one 
at $ \vec{p}=(3.025,0.248)$ rad (setting the lattice spacing $a=1$) (red curves in figure 1)
which is near the $(\pi,0)$ point, or Van Hove singularity region of the BZ, and the second closer to the BZ zone diagonal at $\vec{p}=(1.629,0.822)$ rad (blue curves in figure 1). \\
\\
Proximity to the region of the BZ contributing to the Van Hove singularity in the density of 
states strongly enhances the magnitude of  $\Sigma(\vec{p}, \omega)$. 
In calculations at other Fermi surface momenta (not depicted in the figures in the present paper) in between those shown in figure 1, the magnitude of $\Sigma(\vec{p}, \omega)$ evolves monotonically between the two momentum points displayed in the figure.
The anisotropy of the electronic self energy
$\Sigma(\vec{p}, \omega)$ with Fermi surface momentum $\vec{p}$ shown in figure 1 is due solely to the 
tight binding bandstructure $\varepsilon_{q-p}$ in equation (4). 

\begin{figure}
\begin{center}
\includegraphics[scale=1.0]{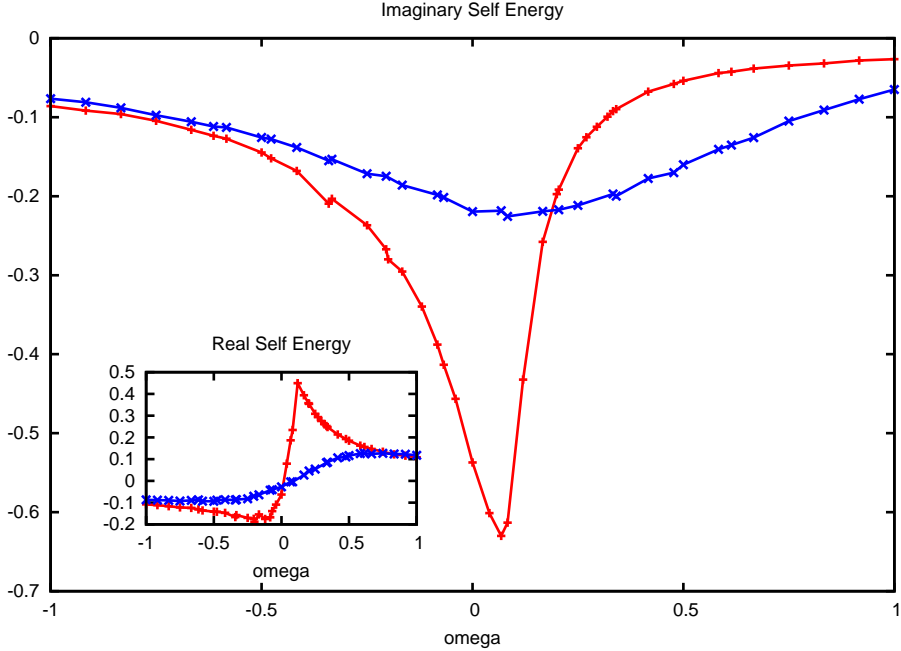}
\caption{Re$\Sigma(\vec{p}, \omega)$ (inset) and Im$\Sigma(\vec{p}, \omega)$ (main figure) 
in units of $t$ for two Fermi surface 
momenta: $ \vec{p}=(3.025,0.248)$ (red curves) and $\vec{p}=(1.629,0.822)$ (blue curves). The vertical
axes are in units of the hopping parameter $t$, and the horizontal axes are $\omega$ in units of $t$. $\mu=-1.3t$ for this case. } 
\end{center}
\end{figure}

The electronic spectral weight $A( \vec{p}, \omega )$ corresponding to the two 
Fermi surface points of figure 1 is shown in figure 2. The enhanced self energy at
$\vec{p}=(3.025,0.248)$  results in the onset of a pseudogap in  $A( \vec{p}, \omega )$,
a feature which is not present for the other Fermi surface point. 
The results shown in figures 1 and 2 are an illustration of the {\em hot spot} nature of
the Van Hove region of the BZ.

\begin{figure}
\begin{center}
\includegraphics[scale=1.0]{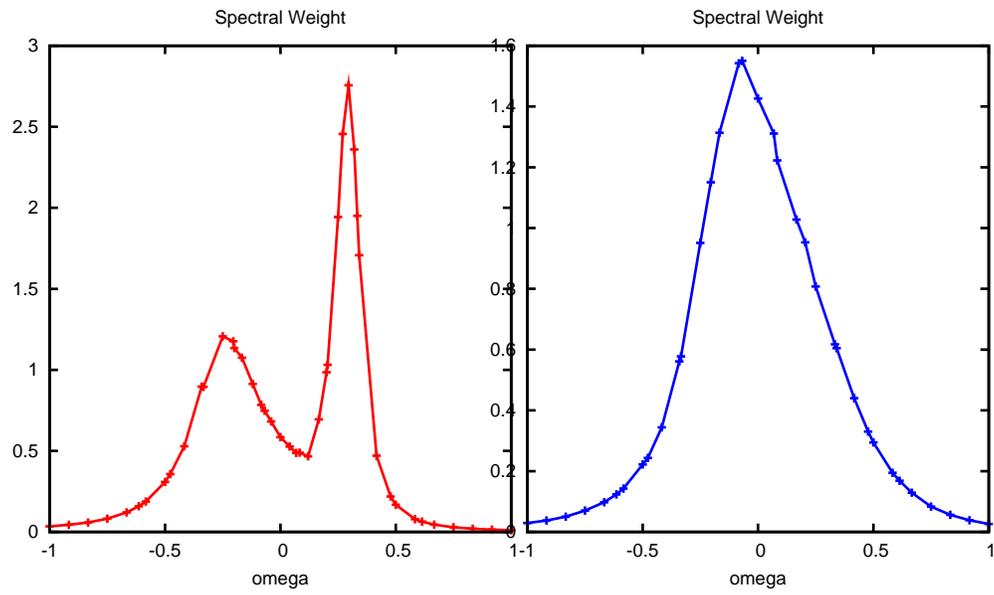}
\caption{The spectral weight $A( \vec{p}, \omega )$ for the two Fermi surface momenta shown in figure 1:
$ \vec{p}=(3.025,0.248)$ (red curves) and $\vec{p}=(1.629,0.822)$ (blue curves)}
\end{center}
\end{figure}

The lack of symmetry in figures 1 and 2 
about $\omega=0$ is a consequence of the tight binding bandstructure. This is illustrated with a choice of $t^{'}=0$ and $\mu=0.1t$ in equation(3) for $\varepsilon_{k}$ which places the Fermi energy near the middle of a symmetric band. The Im$\Sigma(\vec{p}, \omega)$ and $A( \vec{p}, \omega )$, which are now almost symmetrical about $\omega=0$,  are shown in figure 3.

\begin{figure}
\begin{center}
\includegraphics[scale=1.0]{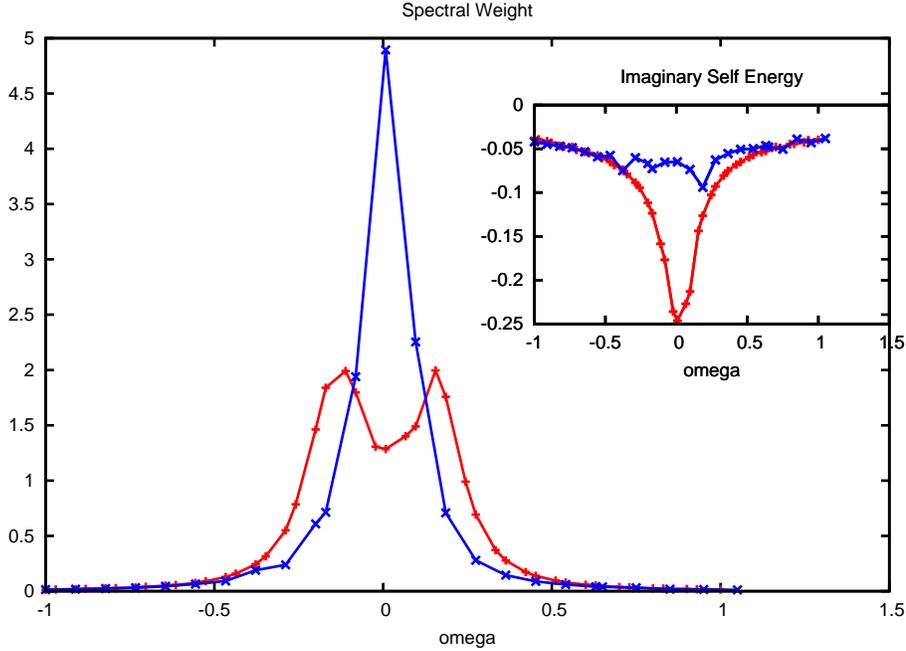}
\caption{Im$\Sigma(\vec{p}, \omega)$ (inset) in units of $t$ and $A( \vec{p}, \omega )$ 
for $t^{'}=0$, $\mu=0.1t$, $\vec{p}=(3.025,0.3385)$ (red curves) and 
$\vec{p}=(1.5126,1.6792)$ (blue curves), $T_{C}=0.15t$, $c=-1.75315t$ and $T=0.17t$. } 
\end{center}
\end{figure}

The results of figures 1 and 2 are calculated with $\mu=-1.3t$, which is close to the
energy of the Van Hove singularity at $-1.4t$ in the tight binding band density of states.
To investigate the effect of a different choice of $\mu$, results for $\Sigma(\vec{p}, \omega)$ 
and $A( \vec{p}, \omega )$
with $\mu =-0.9t$ are shown in figure 4 and 5 for two Fermi surface points: one at $\vec{p}=
(1.629,1.061)$ rad (blue curves), and a second near the $(\pi,0)$ point at $\vec{p}=(3.025,0.552)$ rad.
(red curves).
In this case, the superconducting coupling constant $c=-1.88257t$ in equation (1), resulting in
a $T_{C}=0.2t$.

\begin{figure}
\begin{center}
\includegraphics[scale=1.0]{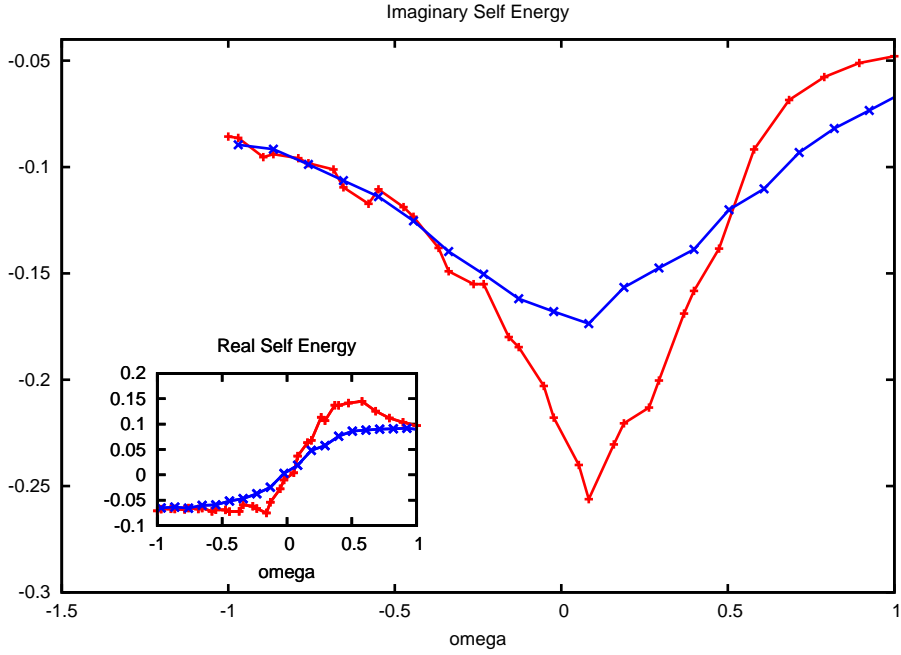}
\caption{Re$\Sigma(\vec{p}, \omega)$ (inset) and Im$\Sigma(\vec{p}, \omega)$ (main figure) 
in units of $t$ for two Fermi surface 
momenta: $ \vec{p}=(3.025,0.552)$ (red curves) and $\vec{p}=(1.629,1.061)$ (blue curves). The vertical
axes are in units of the hopping parameter $t$, and the horizontal axes are $\omega$ in units of $t$. $\mu=-0.9t$ for this case.}
\end{center}
\end{figure}
  
Figure 4 shows a significant reduction in the magnitude of $\Sigma(\vec{p}, \omega)$ compared to the results of figure 1 for
both Fermi surface momenta in the $\mu=-0.9t$ case. Furthermore, while their is a small peak visible in 
the Im$\Sigma(\vec{p}, \omega)$ in figure 4 for $\vec{p}=(3.025,0.552)$ (red curve), which would evolve into the larger peak in the
$\mu=-1.3t$  results of figure 1 as $\mu$ is adjusted to $-1.3t$, the overall difference in magnitude 
of the two Fermi surface $\Sigma(\vec{p}, \omega)$ in figure 4 
is much smaller for $\mu=-0.9t$ compared with the case $\mu=-1.3t$. In other words, the Fermi surface anisotropy in the magnitude of the self energy is significantly weaker in the $\mu=-0.9t$ case.

\begin{figure}
\begin{center}
\includegraphics[scale=1.0]{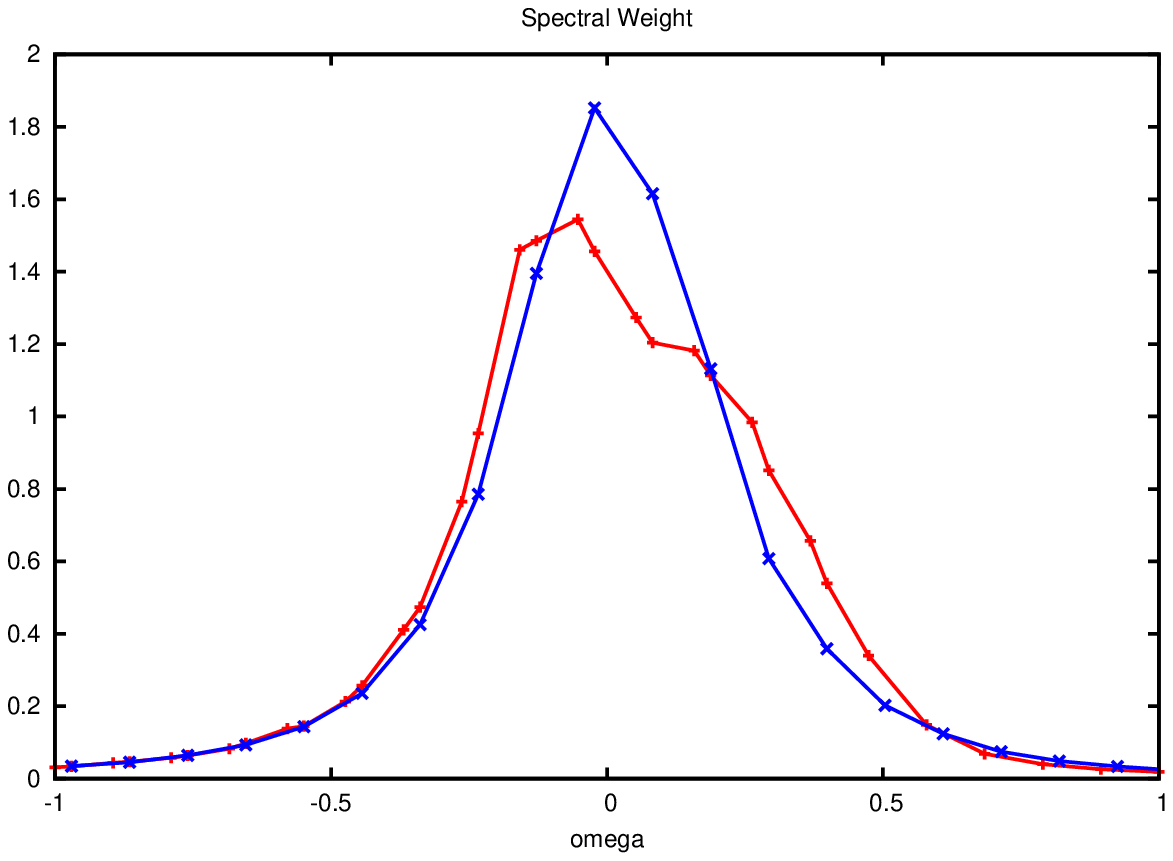}
\caption{The spectral weight $A( \vec{p}, \omega )$ for the two Fermi surface momenta shown in figure 4:
$ \vec{p}=(3.025,0.552)$ (red curve) and $\vec{p}=(1.629,1.061)$ (blue curve)}
\end{center}
\end{figure}

The spectral weights $A( \vec{p}, \omega )$ for 
$\mu=-0.9t$ are shown in figure 5.   
These results show that choosing a chemical potential $\mu$ further away from 
the position of the Van Hove singularity in the
density of states reduces the magnitude of, and the Fermi surface anisotropy in
the fluctuation self energy $\Sigma(\vec{p}, \omega)$, along with eliminating the pseudogap
in $A( \vec{p}, \omega )$.\\
\\
\\
{\bf Conclusion}\\
\\
The role of a tight binding bandstructure in determining the self energy $\Sigma(\vec{p}, \omega)$ due to superconducting fluctuations, and the spectral weight $A( \vec{p}, \omega )$, has been studied. Certain choices of the chemical potential $\mu$ lead to strong anisotropy in these quantities on the Fermi surface, with the onset of an anisotropic pseudogap occuring in the spectral weight.\\
\\
The results may be of use in understanding the Fermi surface anisotropy of the normal state pseudogap 
in high $T_{C}$ cuprates. This is non-zero at the $(\pi,0)$ point, decreasing to zero along
an arc of the Fermi surface about the BZ zone diagonal.\\
\\
Superconducting fluctuations are unlikely to be the sole origin of the observed pseudogap in the cuprates since
the pseudogap increases in magnitude while the superconducting phase disappears as the samples become increasingly underdoped.
One possible scenario for the high $T_{C}$ pseudogap
is that it results from a combination of superconducting fluctuations of the type studied in this work, and a 
separate effect which manifests itself predominantly in the underdoped phase, competing with superconductivity. 
The latter is possibly due to the opening of a correlation gap associated with
the onset of the insulating antiferromagnetic state at low doping \cite{Kampf}.\\ 
\\
A pseudogap due to superconducting fluctuations may be present for $T>T_{C}$  in
the optimal to overdoped region of the cuprate phase diagram. As the doping level is adjusted deeper into the underdoped phase, a transition, occurs over a doping range below optimal doping, to a different pseudogap, which competes with superconductivity.\\
\\    
The similarity in the Fermi surface momentum dependence of the two pseudogaps, originating from two completely
different mechanisms, could then be a consequence of what both mechanisms have in common: the underlying cuprate tight binding bandstructure, and the accompanying {\em hot spot} physics which plays an important role in determining 
self energy effects and resulting spectral weight in both cases.\\
\\
\\

\end{document}